\documentclass{aa}  
\usepackage{natbib}
\bibpunct{(}{)}{;}{a}{}{,} 
\usepackage{graphicx}
\usepackage{capt-of}
\usepackage{txfonts}
\usepackage[version=4]{mhchem}
\newcommand\Tstrut{\rule{0pt}{2.5ex}} 
\usepackage{hyperref}

\begin{document}

   \title{Evidence for chromium hydride in the atmosphere of hot Jupiter WASP-31b}

   \author{Marrick Braam
          \inst{1}\fnmsep\inst{2}\fnmsep\inst{3}\thanks{\email{\href{mailto:mbraam@ed.ac.uk}{mbraam@ed.ac.uk}}}
          \and
          Floris F. S. van der Tak \inst{1}\fnmsep\inst{4}
          \and
          Katy L. Chubb \inst{5}
          \and
          Michiel Min \inst{5}
          }

   \institute{Kapteyn Astronomical Institute, University of Groningen, 
   Landleven 12, 9747 AD Groningen, The Netherlands
         \and 
            School of GeoSciences, University of Edinburgh, 
            King’s Buildings, Edinburgh EH9 3FF, UK
         \and 
            Centre for Exoplanet Science, University of Edinburgh, 
            Edinburgh EH9 3FD, UK
         \and 
             SRON Netherlands Institute for Space Research, 
             Landleven 12, 9747 AD Groningen, The Netherlands
         \and
             SRON Netherlands Institute for Space Research, 
             Sorbonnelaan 2, 3584 CA Utrecht, The Netherlands
             }

   \date{Received September 23, 2020; accepted November 24, 2020}

  \abstract
   {The characterisation of exoplanet atmospheres has shown a wide
   diversity of compositions. Hot Jupiters have the appropriate temperatures to host metallic compounds, which should be detectable through transmission spectroscopy.}
   {We aim to detect exotic species in the transmission spectra of hot Jupiters, specifically WASP-31b, by testing a variety of chemical species to explain the spectrum.}
   {We conduct a re-analysis of publicly available transmission
   data of WASP-31b using the Bayesian retrieval framework \textsc{TauREx II}. We retrieve various 
   combinations of the opacities of 25 atomic and molecular species to 
   determine the minimum set that is needed to fit the observed spectrum.}
   {We report evidence for the spectroscopic signatures of chromium hydride (CrH), H$_2$O, and K in WASP-31b. Compared to a flat model without any signatures, a CrH-only model is preferred with a statistical significance of ${\sim}3.9\sigma$. A model consisting of both CrH and H$_2$O is found with ${\sim}2.6$ and ${\sim}3\sigma$ confidence over a CrH-only model and an H$_2$O-only model, respectively. Furthermore, weak evidence for the addition of K is found at ${\sim}2.2\sigma$ over the H$_2$O+CrH model, although the fidelity of the data point associated with this signature was questioned in earlier studies. Finally, the inclusion of  collision-induced absorption and a Rayleigh scattering slope (indicating the presence of aerosols) is found with ${\sim}3.5\sigma$ confidence over the flat model. This analysis presents the evidence for signatures of CrH in a hot Jupiter atmosphere. At a retrieved temperature of $1481^{+264}_{-355}\,\mathrm{K}$, the atmosphere of WASP-31b is hot enough to host gaseous Cr-bearing species, and the retrieved abundances agree well with predictions from thermal equilibrium chemistry. Furthermore, the retrieved abundance of CrH agrees with the abundance in an L-type brown dwarf atmosphere. However, additional retrievals using VLT FORS2 data lead to a non-detection of CrH. Future observations with JWST have the potential to confirm the detection and/or discover other CrH features.}
   {}

   \keywords{Planets and satellites: atmospheres --
             Planets and satellites: individual: WASP-31b  --
             Techniques: spectroscopic
               }

   \maketitle
%

\section{Introduction}
Soon after the discoveries of the first exoplanets \citep{wolszczanfrail1992, mayorqueloz1995}, their atmospheres became a curiosity (e.g. \citealt{seagersasselov2000}). Nowadays, the confirmed number of exoplanets has exceeded 4000\footnote{Based on data in the \href{https://exoplanetarchive.ipac.caltech.edu/index.html}{NASA Exoplanet Archive}.} and this number is expected to increase significantly over the coming years. Even more remarkable than the large number of discoveries itself is the wide parameter space in which these planets are being found: Equilibrium temperatures range from ${\sim}100{-}4050$ K and masses and radii are continuously found within ranges of $0.1{-}10^{4}\,\mathrm{M}_\oplus$ and $0.3{-}25\,\mathrm{R}_\oplus$. Naturally, an enormous diversity in exoplanet atmospheres can be expected.

Currently, the main method for characterising these exoplanet atmospheres is through transmission spectroscopy (e.g. \citealt{crossfield2015}). A transmission spectrum measures the dip in the stellar light when a planet transits in front of its host star. If the planet has an atmosphere, the opacity and, consequently, the apparent planet size change as a function of wavelength. The atmospheric composition and physical structure can be inferred from these variations with wavelength \citep{seagersasselov2000}. Using the Space Telescope Imaging Spectrograph (STIS) on the Hubble Space Telescope (HST), the first detection of an exoplanet atmosphere was the discovery of the sodium (Na) doublet during a transit of the hot Jupiter HD 209458b \citep{charbonneauetal2002}. Since then, evidence for the features of a variety of other chemical species has been reported, such as H$_2$O, CH$_4$, CO, CO$_2$, and K (see \citet{madhusudhan2019} for an overview). Furthermore, the existence of metallic compounds such as TiO, VO, and AlO has been found on several planets (e.g. \citealt{sedaghatietal2017, evansetal2017, chubb2020w43b}). 

The search for absorption signatures of metallic compounds is inspired by their detections in brown dwarfs  (e.g. \citealt{kirkpatricketal1999, kirkpatrick2005, loddersfegley2006}), and they are also predicted to be important species in the temperature ranges of hot exoplanets (e.g. \citealt{burrowssharp1999, woitkehelling2018}). Amongst these metallic compounds, chromium hydride (CrH) and iron hydride (FeH) are relevant in the brown dwarf classification scheme, notably in specifying the transition from L to T dwarfs \citep{kirkpatrick2005}. The detections of atomic metal species in ultra-hot Jupiters, such as Cr I, Fe I, Mg I, Na I, Ti I, and V I  \citep{hoeijmakersetal2018, hoeijmakersetal2019, benyamietal2020}, suggest that the hydrides CrH and FeH can also be expected in the atmospheres of hot exoplanets. Tentative detections of FeH have been reported for four planets: WASP-62b \citep{skafetal2020}, WASP-79b \citep{skotzenetal2020, skafetal2020}, WASP-121b \citep{evans2016}, and WASP-127b \citep{skafetal2020}, whereas \citet{kesselietal2020} did not find statistically significant detections for 12 planets using high dispersion transmission spectroscopy. Furthermore, evidence for the presence of metal hydrides in the exo-Neptune HAT-P-26b was found by \citet{macdonald2019}, who identified three possible candidates to explain these features in the optical part of the transmission spectrum: TiH, CrH, or ScH. Found as part of the Wide Angle Search for Planets \citep{pollacco2006}, WASP-31b is thought to be in the right temperature range to host metal hydrides.

WASP-31b, which is in orbit around an F-type star,  was discovered by \citet{anderson2011}. The planet has a mass of $0.478\,\mathrm{M_J}$ and a radius of $1.549\,\mathrm{R_J}$, making it one of the lowest density exoplanets known to date. Orbiting at a distance of $0.047\,\mathrm{AU}$ from its host star, it has an equilibrium temperature of $1393\,\mathrm{K}$ (assuming Jupiter's Bond albedo of $0.34$). Its low density (surface gravity) and high temperature lead to a large scale height, making WASP-31b a suitable candidate for atmospheric characterisation using transmission spectroscopy. Its host star has an effective temperature of $6300\pm100\mathrm{K}$ and a metallicity of ${-}0.20\pm0.09\,\mathrm{dex}$. The system age is estimated to be $1^{+3}_{-0.5}\,\mathrm{Gyr}$ \citep{anderson2011}. Using optical to mid-infrared transmission spectra to probe the atmosphere of WASP-31b, \citet{sing2015} found a strong potassium (K) feature as well as evidence for the presence of aerosols, both in the form of clouds (grey scatter) and hazes (Rayleigh scatter). Evidence for a grey cloud deck was also obtained by \citet{barstowetal2017}, whereas other comparative studies found some weak H$_2$O \citep{pinhasetal2019, welbanksetal2019} and NH$_3$ features \citep{macdonaldetal2017, minetal2020}. The existence of the K signature has been called into question by recent observations using the FOcal Reducer and low dispersion Spectrograph 2 (FORS2) and the Ultraviolet and Visual Echelle Spectrograph (UVES) on the Very Large Telescope (VLT) \citep{gibson2017, gibson2019} and the Inamori-Magellan Areal Camera and Spectrograph (IMACS) on the Magellan Baade Telescope \citep{mcgruderetal2020}.

In this study, we conduct a re-analysis of the publicly available transmission spectrum of WASP-31b using the \textsc{TauREx} retrieval framework \citep{waldmann2015}. In Section \ref{sec:methodology}, we describe the observations that were used in this analysis and provide the details of the retrieval setup. The retrieval results are presented in Section \ref{sec:results}. In Section \ref{sec:discussion}, we compare our findings with earlier detections and discuss the physical implications before providing the conclusions in Section \ref{sec:conclusion}.

\section{Methodology}\label{sec:methodology}
\subsection{Observations}
The optical and near-infrared transit light curves of WASP-31b were observed  using HST and then analysed by \citet{sing2015}. Transits were observed using STIS with the G430L and G750L gratings, providing spectral coverage from $0.29$ to $1.027\,\mu \mathrm{m}$ at a resolution of $530{-}1040$. These were supplemented by observations from $1.1$ to $1.7\,\mu \mathrm{m}$ at R${\sim}70$ using the G141 grism of the Wide Field Camera 3 (WFC3). \citet{sing2015} combined their observations with photometric measurements in the $3.6$ and $4.5\,\mu \mathrm{m}$ channels obtained using Spitzer's Infrared Array Camera (IRAC)\footnote{See \href{https://pages.jh.edu/\~dsing3/David\_Sing/Spectral\_Library.html}{https://pages.jh.edu/\(\sim\)dsing3/David\_Sing/Spectral\_Library.html} and \href{https://stellarplanet.org/science/exoplanet-transmission-spectra/}{https://stellarplanet.org/science/exoplanet-transmission-spectra/}.}. 

From the planetary parameters in Table \ref{table:planetparams}, it can be seen that WASP-31b is larger than Jupiter and orbits close to its host star. Furthermore, with only half of Jupiter's mass, the planet is one of the lowest density planets known. We calculated the equilibrium temperature assuming Jupiter's Bond albedo of $0.34$ and a redistribution factor ${f}{=}1$ defining isotropic re-emission; for the lower and upper boundaries, we assumed ${A}{=}0.9,\,{f}{=}1$ and ${A}{=}0.12,\,{f}{=}0.5$, respectively. 

\begin{table*}
\caption{Planetary parameters}             
\label{table:planetparams}  
\centering  
\begin{tabular}{l l l l l l l l}     
\hline \hline   
$\mathrm{R_p(R_J)}$ & $\mathrm{M_p(M_J)}$ & SMA(AU) & $\mathrm{T_{eq}(K)}$\tablefootmark{a} & $\mathrm{T_{range}}$\tablefootmark{a} & $\mathrm{R_*(R_\odot)}$ & $\mathrm{T_*(K)}$ & Reference \vspace{1mm} \\
\hline                    
$1.549{\pm}0.050$ & $0.478{\pm}0.029$ & $0.0466{\pm}0.0004$ & $1393$  & $869{-}1882$ & $1.252{\pm}0.033$ & $6302{\pm}102$ & \citet{anderson2011} \Tstrut    \\
\hline                  
\end{tabular}
\tablefoot{
\tablefoottext{a}{Equilibrium temperatures were calculated under varying assumptions for the Bond albedo and redistribution factor (see text).}
}
\end{table*}

\subsection{TauREx II}
The observed transmission spectra are a complex function of many underlying parameters, and acquiring information about these parameters is known as the inverse, or retrieval, problem. We need to determine, given the planetary spectrum that is observed, what the most likely composition and state of the planetary atmosphere are. The retrieval was conducted using \textsc{TauREx II} \citep{waldmann2015}, which is a Bayesian retrieval framework based on a forward model that computes 1D atmospheric radiative transfer \citep{hollis2013}. The propagation of radiation through an atmosphere is strongly dependent on pressure-temperature structure and composition. \textsc{TauREx} maps the correlations between atmospheric parameters and provides statistical estimates on their values. 

In planetary atmospheres, there are two forms of interaction between stellar radiation and the gases in an atmosphere: scattering and absorption of radiation. For molecular opacities, \textsc{TauREx} relies on the ExoMol \citep{jt631}, HITEMP \citep{HITEMP}, and MoLLIST \citep{MOLLIST} databases. These databases contain line lists of many molecular species, providing their energy levels and transition probabilities, up to high temperatures. This allows us to compute the wavelength-dependent absorption of a particular species as a function of temperature and pressure. The line lists that are used in our analysis are shown in Table \ref{table:linelistdata}. There is a threefold motivation for the choice of chemical species. Firstly, thermal equilibrium chemistry can predict the presence and expected abundances of species as a function of temperature, pressure, and elemental composition \citep{woitkehelling2018}. Assuming initial solar composition, this predicts the main element-bearing species at the temperatures of hot Jupiters to be, for example, H$_2$O and CO for oxygen, CO, CO$_2$, and CH$_4$ for carbon, and TiO for titanium. Secondly, a possible detection requires the species to have signatures in the observed spectral regime. Knowledge of the absorption signatures of metal hydrides and oxides is informed by their detections in brown dwarfs (e.g. \citealt{kirkpatricketal1999, loddersfegley2006, sharpburrows2007}), whereas the prominent signatures of Na (near $0.59\,\mu \mathrm{m}$) and K (near $0.77\,\mu \mathrm{m}$) are identifiable in the optical regime (e.g. \citealt{charbonneauetal2002, welbanksetal2019}). Most of the nitrogen is expected to be present in N$_2$. Being a homonuclear diatomic molecule, N$_2$ has no prominent signatures in the infrared. NH$_3$, especially prominent at cool temperatures, is the next main nitrogen-bearing species and does have signatures, motivating its inclusion. Lastly, species such as C$_2$H$_2$, HCN, and OH are expected to be related to photochemical processes (e.g. \citealt{linetal2010, kawashima2019}). 

\begin{table*}
\caption{Atomic and molecular data used in this analysis.}  
\label{table:linelistdata}      
\centering          
\begin{tabular}{l l l l}     
\hline\hline                   
Molecule   & Wavelength range & Number of lines   & Database/reference \vspace{1mm} \\ 
\hline          
AlH         & $0.37{-}100\mu$m  & 36,000 & ExoMol: \citet{alhopacity}    \\
AlO         & $0.29{-}100\mu$m  & 4,945,580   & ExoMol: \citet{ExoMol_AlO}     \\
C$_2$H$_2$  & $1.00{-}100\mu$m  &  4,347,381,911  & ExoMol: \citet{ExoMol_C2H2}   \\
C$_2$H$_4$  & $1.41{-}100\mu$m  & 49,841,085,051  & ExoMol: \citet{jt729}     \\
CaH         & $0.45{-}100\mu$m  & 19,095 & MoLLIST: \citet{11LiHaRa.CaH, MOLLIST} \\
CH$_4$      & $0.83{-}100\mu$m  & 34,170,582,862  &  ExoMol: \citet{ExoMol_CH4}      \\
CN          & $0.23{-}100\mu$m  & 195,120  & MoLLIST: \citet{14BrRaWe.CN}     \\
CO          & $0.45{-}100\mu$m  & 752,976  & \citet{15LiGoRo.CO}     \\
CO$_2$      & $1.04{-}100\mu$m & 11,167,618 & HITEMP: \citet{HITEMP}          \\
CP          & $0.67{-}100\mu$m  & 28,752  & MoLLIST: \citet{14RaBrWe.CP}     \\
CrH         & $0.67{-}100\mu$m  & 13,824 & MoLLIST: \citet{02BuRaBe.CrH, MOLLIST} \\
FeH         & $0.67{-}100\mu$m  & 93,040 & MoLLIST: \citet{10WEReSe.FeH}    \\
H$_2$CO     & $0.99{-}100\mu$m  & 12,688,112,669  & ExoMol: \citet{jt597}     \\
H$_2$O      & $0.24{-}100\mu$m  & 5,745,071,340 & ExoMol: \citet{ExoMol_H2O_POK}  \\
HCN         & $0.56{-}100\mu$m  & 34,418,408  & ExoMol: \citet{ExoMol_HCN}      \\
K           & $0.29{-}100\mu$m  & 186 & NIST: \citet{NISTWebsite, 16AlSpKi.broad}     \\
MgH         & $0.34{-}100\mu$m  & 30,896 & MoLLIST: \citet{13GhShBe.MgH}        \\
MgO         & $0.27{-}100\mu$m  & 72,833,173  & ExoMol: \citet{ExoMol_MgO}     \\
Na          & $0.24{-}100\mu$m  & 523 & NIST: \citet{NISTWebsite, 19AlSpLe.broad}     \\
NH$_3$      & $0.43{-}100\mu$m  & 16,941,637,250 & ExoMol: \citet{ExoMol_NH3}      \\
OH          & $0.23{-}100\mu$m  & 54,276 & MoLLIST: \citet{18YoBeHo.OH}     \\
ScH         & $0.63{-}100\mu$m  & 1,152,826  & LYT: \citet{jt599}         \\
TiH         & $0.42{-}100\mu$m  & 199,072  &  MoLLIST: \citet{05BuDuBa.TiH}        \\
TiO         & $0.33{-}100\mu$m  & 59,324,532 & ExoMol: \citet{ExoMol_TiO}      \\
VO          & $0.29{-}100\mu$m  & 277,131,624 & ExoMol: \citet{ExoMol_VO}       \\
\hline                  
\end{tabular}
\end{table*}

The data in Table \ref{table:linelistdata} were converted into cross-sections and k-tables  by \citet{19Chubbetal} in order to feed them into \textsc{TauREx}  as part of the ExoMolOP database \footnote{\url{http://www.exomol.com/data/data-types/opacity/}}. In this study, we have used the k-tables with $R{=}\frac{\Delta \lambda}{\lambda}{=}300$. A k-table provides the absorption coefficient as a function of wavelength for a certain temperature and pressure.

Furthermore, \textsc{TauREx} includes the continuum opacity caused by collision-induced absorption (CIA) of H$_2$-H$_2$ and H$_2$-He pairs and a parametrisation for the opacity caused by particle scattering. This represents the interaction between radiation and aerosols, thus quantifying the influence of clouds and hazes. For atmospheric particles that are small relative to the wavelength of the incoming light, there is a strong $\mathrm{\lambda}^{-4}$ dependence of Rayleigh scattering. The opacity due to Rayleigh scattering is based on pre-computed cross-sections \citep{hollis2013}. Besides that, an optically thick grey cloud cover would lead to a flat opacity as a function of $\mathrm{\lambda}$ and is modelled via the cloud top-pressure $\mathrm{P_{cl}}$. \textsc{TauREx} also contains a more complex cloud model that parametrises the opacity due to the scattering of light by spherical particles, following the Mie theory \citep{lee2013}. This parametrisation was tried but not found to be significant.

\subsection{Retrieval}
\textsc{TauREx} searches the multi-dimensional parameter space for solutions through the \textsc{MultiNest} algorithm \citep{ferozhobson2008, ferozetal2009, ferozetal2013}. As an output, \textsc{MultiNest} provides the global log-evidence, or simply the Bayesian evidence, which tests the adequacy of the model itself and can be used to compare models of varying complexity. In this comparison, Occam's razor is applied: Adding a factor of complexity to an atmospheric model is only appropriate when this inclusion gives a significantly better fit to the data. When comparing two models, $\mathcal{M}_2$ having an extra atmospheric parameter and thus more complexity than $\mathcal{M}_1$, their Bayesian evidence can be used to calculate the ratio of the model probabilities, or the Bayes factor \citep{kassraftery1995, waldmann2015},
\begin{equation}\label{bayesfactor}
    \mathcal{B}_{21}=\frac{\mathrm{E_2}}{\mathrm{E_1}}
,\end{equation}
or to define the detection significance (DS),
\begin{equation}\label{eqn:bayesfactor}
    \mathrm{DS} = \ln(\mathcal{B}_{21})= \ln(\mathrm{E_2}) - \ln(\mathrm{E_1})
.\end{equation}
Table 3 shows the empirically calibrated Jeffreys' scale from \citet{trotta2008}, which we used to quantify the preference for an additional atmospheric parameter: A DS greater than one provides evidence in favour of the more complex model. We refer to this as the 'detection significance' since more complexity is usually represented by the addition of a particular chemical species to our model.

\begin{table}
\caption{Empirically calibrated Jeffreys' scale \citep{jeffreys1998} with translation to frequentist values specifying the odds in favour of the more complex model, adapted from \citet{trotta2008}.}             
\label{table:jeffreysscale}      
\centering          
\begin{tabular}{l l l l} 
\hline\hline                   
    $\mathrm{DS}$ & p-value & $\sigma$ & Category \vspace{1mm} \\ \hline
    1.0     & 0.04          & 2.1   &  'Weak' at best  \Tstrut    \\
    2.5     & 0.006         & 2.7   &  'Moderate' at best  \\ 
    5.0     & 0.0003        & 3.6   &  'Strong' at best    \\ 
    11.0    & 6$\times10^{-7}$   & 5.0   & 'Very strong' \\
\hline                  
\end{tabular}
\end{table}

\subsection{General setup}
The atmospheric models consist of 100 isothermal layers with pressures ranging from $10^{6}$ to $10^{-5}$ Pa. We assumed a hydrogen-dominated atmosphere with a Jupiter-like $\mathrm{He}/\mathrm{H}_2{=}0.157$ and used the prior values and temperature boundaries from Table \ref{table:planetparams}. The planetary radius was fitted within ranges of $0.1\,\mathrm{R_p}$ around the prior value, and the retrieved presence of a grey cloud cover was allowed in the full pressure range. Furthermore, the chemical abundances were retrieved with volume mixing ratios (VMRs) or abundances between $10^{-10}$ and $10^{-1}$. From the retrieved abundances, the atmospheric molecular weight was then calculated. In the end, up to 28 free parameters can thus be retrieved in the procedure. However, given the limited number of data points and spectral coverage, only a small fraction of these parameters will statistically be required. Together with the feasibility of quantifying the importance of individual molecules, this is one of the main reasons for a bottom-up approach.

In this bottom-up approach, the retrieval was first performed assuming the simplest atmospheric forward model consisting of three free parameters ($\mathrm{R_p}$, $\mathrm{T}$, and $\mathrm{P_{cl}}$), which is equivalent to an atmosphere completely lacking spectral signatures. Afterwards, retrievals were done by adding parameters to the atmospheric model in the form of the abundance of a chemical species. The first stage is to compare the models with a single chemical species to the flat model ($\mathrm{R_p}$, $\mathrm{T}$, and $\mathrm{P_{cl}}$ only), using the DS (see Eq. \ref{eqn:bayesfactor}). As opposed to the flat model, the opacities caused by Rayleigh scattering and CIA are from now on also included. If the addition of a chemical species leads to an improved fit to the data, this results in stronger evidence, and the significance of such a detection is specified by Jeffreys' scale in Table \ref{table:jeffreysscale}. As a second stage, we tested the inclusion of H$_2$O plus another species, mainly because H$_2$O has a prominent absorption feature in the relatively well-covered near-infrared (e.g. \citealt{sing2016, tsiaras2018}). Evidence levels from these models can be compared to the models containing a single species as well as to the flat model. Lastly, the model with the strongest evidence was expanded by adding the absorption features of the alkali metals Na and K in order to quantify their possible presence.

\section{Results}\label{sec:results}
\begin{figure*}
\centering
\includegraphics[width=1.0\linewidth]{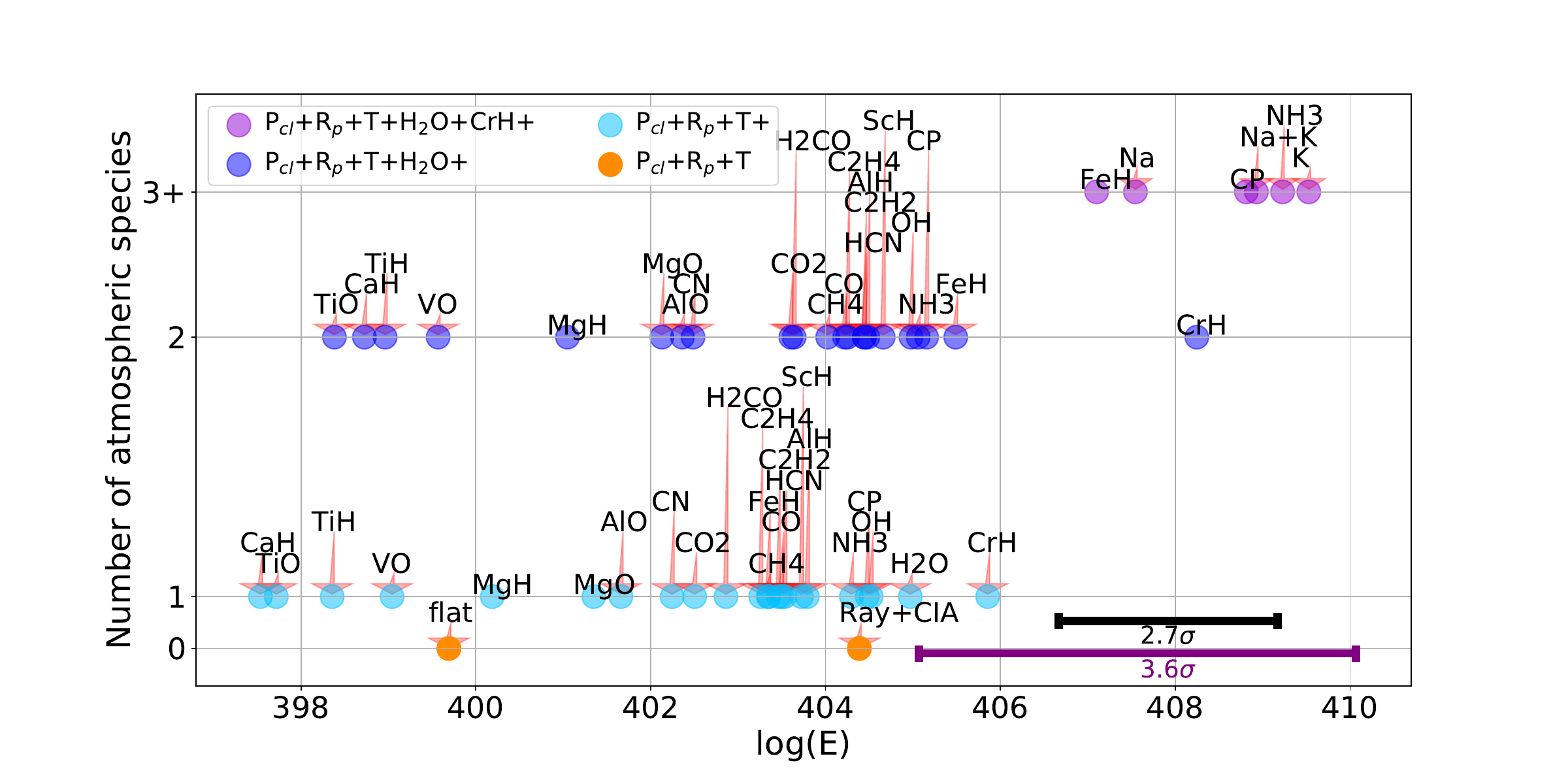}
\caption{Model comparison for WASP-31b using Bayesian evidence for different atmospheric models. The flat model is shown as the orange dot, whereas cyan dots indicate higher complexity in the form of a chemical species, as labelled. One stage higher, blue dots represent a model that includes H$_2$O and an additional parameter. For the final stage, the strongest evidence model from lower complexities is complemented by K and potentially Na, as shown by the violet dots. In this final stage, we also tested the inclusion of some of the more likely species of lower levels. The horizontal scale bars indicate the statistical preference for a more complex model, based on Table \ref{table:jeffreysscale}.}
\label{fig:evw31b}
\end{figure*}

Following the bottom-up approach for WASP-31b, its spectrum was retrieved assuming 53 different atmospheric models. The resulting evidence for each of these models can be seen in Figure \ref{fig:evw31b}, and the exact values for the evidence and DSs (see Eq. \ref{eqn:bayesfactor}) are shown in Table \ref{table:wasp31bresults}. Starting from the lower left of Figure \ref{fig:evw31b}, it can be seen that the flat model without any signatures (only $\mathrm{R_p}$, T, and $\mathrm{P_{cl}}$; represented by an orange dot) leads to a Bayesian evidence level of $399.69$. The inclusion of Rayleigh scattering and CIA is labelled as 'Ray+CIA' and is detected with a confidence level of ${\sim}3.5\sigma$ over the flat model. Except for the 'flat model', all atmospheric models contain the opacity caused by Rayleigh scattering and CIA.

An increase in the complexity, by adding a single atmospheric species on top of the opacity from Rayleigh scattering and CIA, gives the models that are shown as the cyan dots and which are specified by their accompanying labels. It can be seen that the addition of only a small selection of chemical species leads to an increase in evidence levels. The decrease in evidence seen for several species (e.g. CaH and TiO) is caused by our choice of the lower boundary for the abundances in the retrievals. For example, TiO at $\log(\mathrm{X_{TiO}}){=}{-}10$ would still cause absorption features. A negative DS then means that the abundance of the added species is lower than the retrieval boundary, signifying the abundance at which features are no longer seen. 

At this stage, the strongest preference is found for the inclusion of either CrH or H$_2$O, with DSs over the flat model of $6.16$ and $5.28$, respectively. Following Jeffreys' scale (see Table \ref{table:jeffreysscale}), this corresponds to confidence levels of ${\sim}3.9\sigma$ and ${\sim}3.7\sigma$, respectively. Compared to the model containing Rayleigh scattering and CIA, the CrH signature in this single-molecule model is detected at ${\sim}2.3\sigma$. Ascending one stage in complexity, the blue dots show the models consisting of H$_2$O and another species. It can be seen that the combined inclusion of both H$_2$O and CrH is preferred, with a DS of $8.56$ (or ${\sim}4.4\sigma$ confidence) over the flat model, or $3.86$ (${\sim}3.2\sigma$) over a model of only Rayleigh scattering and CIA. Compared to a CrH- or H$_2$O-only model, a model containing both species corresponds to confidence levels of ${\sim}2.6\sigma$ and ${\sim}3.0\sigma$, respectively. 

The final stage is given by the violet dots and represents the addition of further atmospheric species to the best models of previous stages, in this case the one containing H$_2$O and CrH. Regarding the alkali metals, the inclusion of K in the WASP-31b atmospheric model is preferred, with a DS of 1.28 as compared to the H$_2$O+CrH model, whereas the model that includes both Na and K leads to a DS of 0.68. Hence, statistical evidence is only found for the absorption signature of K. The K detection corresponds to weak evidence at a confidence level of ${\sim}2.2\sigma$ over the H$_2$O$+$CrH model. The fact that the detection of K is mainly based on a single strong absorption peak can explain this weak evidence since the signature is covered by just a single data point. Naturally, providing a better fit to only one out of 63 data points may correspond to such a weak increase in Bayesian evidence. In this final stage, we also examined the individual additions of FeH, CP, and NH$_3$ to the H$_2$O$+$CrH model since these species correspond to the highest evidence levels in the previous stage, which included two atmospheric species. Compared to the H$_2$O$+$CrH model, the addition of NH$_3$ or CP leads to a small DS ($0.98$ and $0.56$, respectively) and the addition of FeH leads to a negative DS of ${-}1.15$. None of these are significant according to the Jeffreys' scale.

\begin{table}
\caption{Resulting Bayesian evidence levels and DSs for WASP-31b. Except for the flat model, every model includes the opacity caused by Rayleigh scattering and CIA.}             
\label{table:wasp31bresults}      
\centering          
\begin{tabular}{l c c}     
\\
\hline\hline                   
\textbf{Model parameters }& \textbf{log(E) }&\textbf{ DS }\vspace{1mm} \\ \hline 
Flat model     &   399.69  &      \\
Rayleigh + CIA     &   404.39  & 4.70  \vspace{1mm}    \\ \hline
Compared to flat model & & \\ \hline
CaH     &   397.54  & -2.16     \\
TiO     &   397.72  & -1.97     \\
TiH     &   398.36  & -1.34     \\
VO      &   399.04  & -0.65     \\ 
MgH     &   400.19  & 0.50      \\
MgO     &   401.35  & 1.66      \\
AlO     &   401.66  & 1.97      \\
CN      &   402.25  & 2.55      \\  
CO$_2$  &   402.50  & 2.81      \\
H$_2$CO &   402.86  & 3.17      \\
C$_2$H$_4$ & 403.26 & 3.57      \\
FeH     &   403.34  & 3.65      \\
CH$_4$  &   403.35  & 3.66      \\  
C$_2$H$_2$ &   403.46  & 3.77      \\
CO      &   403.50  & 3.80      \\
HCN     &   403.53  & 3.84      \\  
ScH     &   403.72  & 4.03      \\
AlH     &   403.79  & 4.10      \\
NH$_3$  &   404.30  & 4.60      \\
CP      &   404.48  & 4.78      \\
OH      &   404.52  & 4.83      \\
H$_2$O  &   404.97  & 5.28      \\
CrH     &   405.86  & 6.16      \vspace{1mm} \\ \hline
Compared to H$_2$O-only model & & \\ \hline
H$_2$O + TiO    &   398.38  & -6.59      \\
H$_2$O + CaH    &   398.73  & -6.24      \\
H$_2$O + TiH    &   398.96  & -6.00     \\
H$_2$O + VO     &   399.57  & -5.40      \\
H$_2$O + MgH    &   401.05  & -3.92     \\
H$_2$O + MgO    &   402.13  & -2.84     \\
H$_2$O + AlO    &   402.36  & -2.61     \\
H$_2$O + CN     &   402.48  & -2.48     \\
H$_2$O + CO$_2$ &   403.60  & -1.37      \\
H$_2$O + H$_2$CO &  403.64  & -1.33      \\
H$_2$O + CH$_4$ &   404.02  & -0.95      \\
H$_2$O + CO     &   404.22  & -0.75      \\
H$_2$O + C$_2$H$_4$ &   404.25  & -0.72     \\
H$_2$O + HCN    &   404.44  & -0.53      \\
H$_2$O + C$_2$H$_2$ &   404.44  & -0.52      \\
H$_2$O + AlH    &   404.48  & -0.49     \\
H$_2$O + ScH    &   404.66  & -0.31     \\
H$_2$O + OH     &   404.98  & 0.01      \\
H$_2$O + NH$_3$ &   405.06  & 0.09      \\
H$_2$O + CP     &   405.16  & 0.19     \\
H$_2$O + FeH    &   405.49  & 0.52      \\
H$_2$O + CrH    &   408.25  & 3.28      \vspace{1mm} \\ \hline
Compared to H$_2$O+CrH model & & \\ \hline
H$_2$O + CrH + FeH      & 407.10 & -1.15 \\
H$_2$O + CrH + Na       & 407.55 & -0.70 \\
H$_2$O + CrH + CP       & 408.81 & 0.56 \\
H$_2$O + CrH + Na + K   & 408.93 & 0.68 \\
H$_2$O + CrH + NH$_3$   & 409.23 & 0.98 \\
H$_2$O + CrH + K        & 409.53 & 1.28 \\
\hline 
\end{tabular}
\end{table}

\begin{figure}
\includegraphics[width=0.95\linewidth]{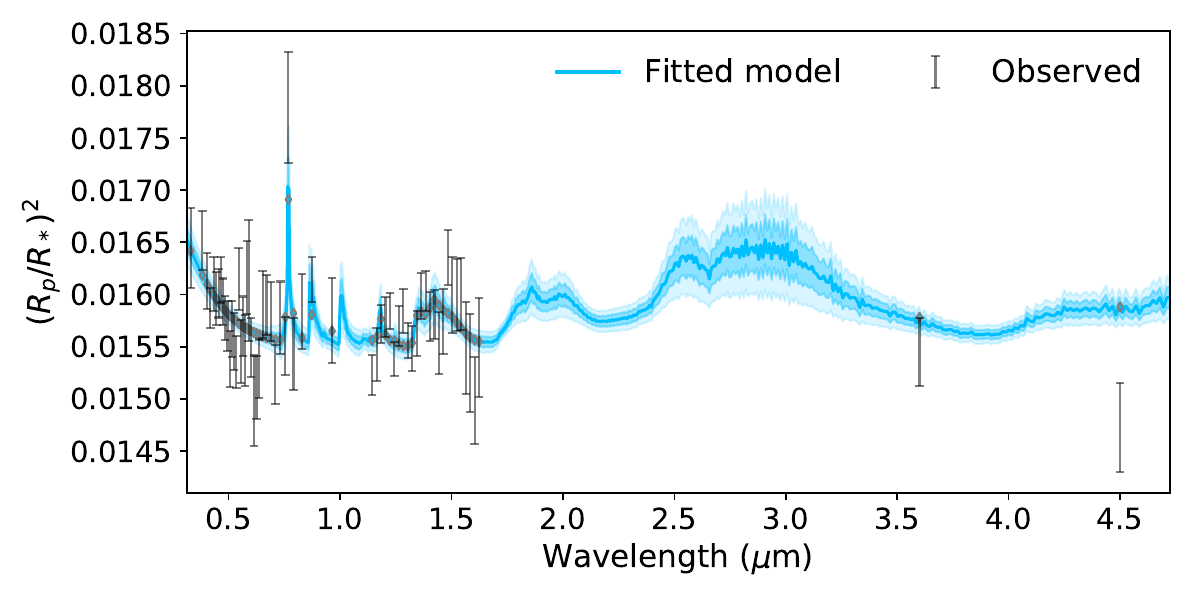}
\includegraphics[width=0.95\linewidth]{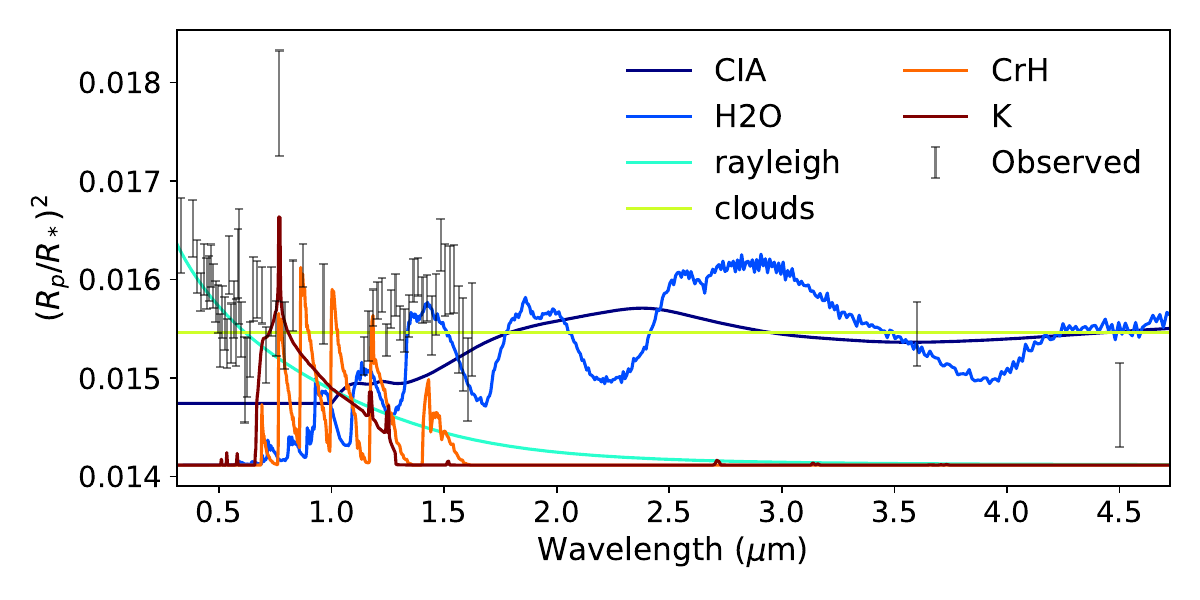}
\caption{\textsc{TauREx} retrieval results for WASP-31b, with the transmission spectrum and the best-fitting atmospheric model (top) and the individual contributions of each molecule to the opacity (bottom). The vertical error bars indicate the observed transit depths, and the different shadings in the upper panel represent $1$ and $2{\sigma}$ regions.}
\label{fig:w31bspec}
\end{figure}

We conclude that out of the models fitted in this study, the spectrum of WASP-31b is best represented by a model that includes H$_2$O, CrH, and K in addition to H$_2$, He, a grey cloud deck, and Rayleigh scattering. This atmospheric model and the observed transmission spectrum of WASP-31b are shown in the upper panel of Figure \ref{fig:w31bspec}. The lower panel shows the individual contributions of the atmospheric constituents to the opacity. The signatures of H$_2$O in the near-infrared (around $1.0, 1.2$, and $1.4\,\mu \mathrm{m}$; in blue) and K in the visible (around $0.77\,\mu \mathrm{m}$; dark red) are easily recognised, whereas the inclusion of CrH leads to the six absorption signatures between $0.7$ and $1.5\,\mu \mathrm{m}$, as shown by the orange line. On top of that, the navy line represents the continuum opacity provided by CIA, and the presence of aerosols results in two distinct signatures: the grey scattering opacity of a low altitude cloud deck and Rayleigh scattering at short wavelengths due to haze.

\begin{figure}
\includegraphics[width=1.0\linewidth]{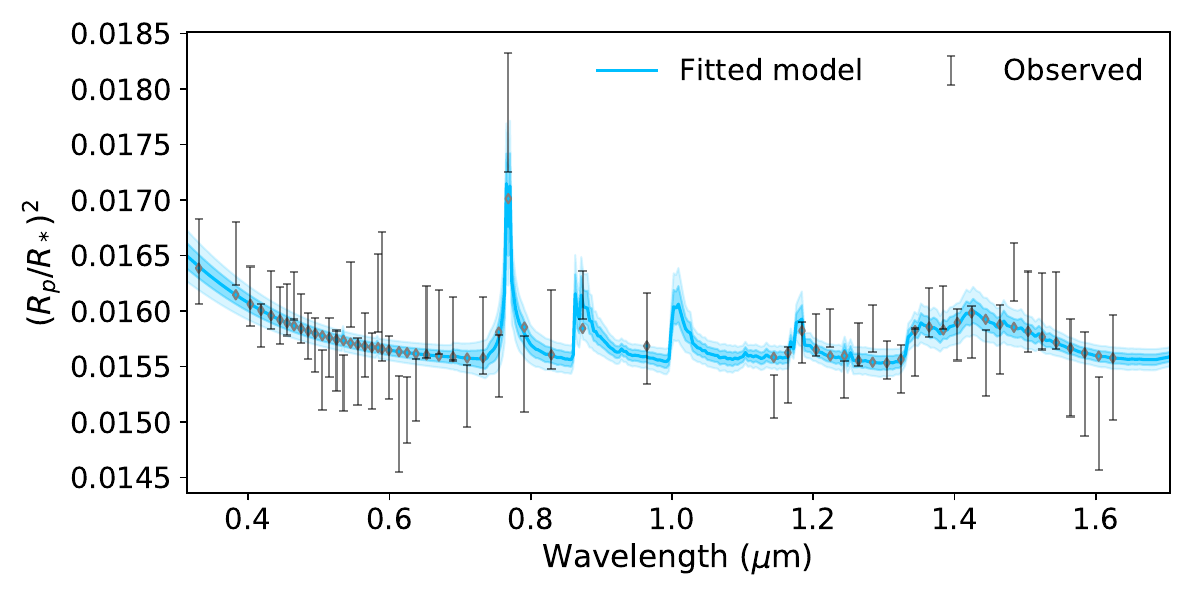}
\includegraphics[width=1.0\linewidth]{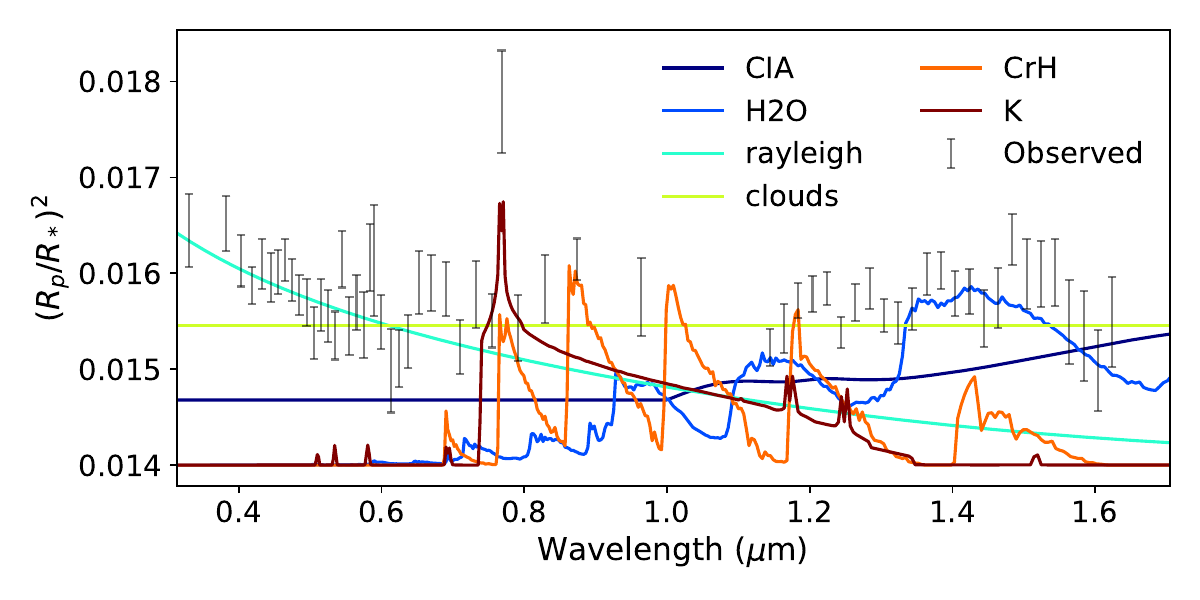}
\caption{\textsc{TauREx} retrieval results for WASP-31b without the Spitzer measurements.}
\label{fig:w31bspecnospitz}
\end{figure}

In Figure \ref{fig:w31bspec}, a discrepancy can be seen between the transit depth resulting from our atmospheric model and the measured transit depth at $4.5\,\mu \mathrm{m}$ as observed by Spitzer. Uncertainties exist in cross-calibrating measurements by different instruments, and the usefulness of Spitzer's broadband photometry in inferring atmospheric compositions has been called into question (see e.g. \citealt{hansenetal2014}). To test our findings, we conducted the same analysis for a spectrum that excludes the Spitzer measurements. The spectrum and its best-fitting atmospheric model can be seen in Figure \ref{fig:w31bspecnospitz}. Excluding the Spitzer measurements leads to a significant preference for the CrH+H$_2$O model, with confidence levels of ${\sim}4.8\sigma$ over a flat model and ${\sim}3.9\sigma$ over a model containing Rayleigh scattering and CIA. Besides that, the retrieved values are consistent with our earlier findings within $1\sigma$. Therefore, we conclude that removing the Spitzer measurements does not change our results significantly, illustrating that mid-infrared coverage is not required to detect CrH. The discrepancy between the data and our model at these wavelengths hints at the influence of species that show spectral activity in these regions (such as CO and CO$_2$).

\section{Discussion}\label{sec:discussion}
Before discussing the retrieved parameters, it is important to emphasise that the resulting atmospheric parameters are based on the models that turned out to be the best fit to the spectral data of WASP-31b. A bottom-up approach is valuable in inferring the presence of chemical species in an atmosphere but may lead to biases in the derived constraints on retrieved parameters. Excluding a particular chemical species from the atmospheric model means that the spectroscopic signature of the species has not been detected on the basis of statistics. This does not necessarily mean that a chemical species is completely absent from the atmosphere that is probed. Instead, the signatures of a species can fall outside of the observed spectral range, be too weak to be detected, or be affected by an overlap with other spectral signatures. The omission of a particular species can then influence the retrieval outcomes since its signatures (even if they are statistically insignificant) have to be explained by the absorption of other species. This may result in unreasonably tight constraints as well as unrealistic values for retrieved abundances. In this way, the retrievals may introduce biases in, for example, the abundances of species that are included in the model. The retrieved parameters for our best-fitting model are shown in Table \ref{table:retrievedparams}, and the extended posterior distributions can be found in Figure \ref{fig:wasp31bpost} of the appendix. To test whether the omission of atmospheric species leads to biases in the retrieved parameters, we retrieved the same spectrum assuming a model with a variety of opacity sources (H$_2$O, CrH, K, CO$_2$, CO, CH$_4$, NH$_3$, and Na). This retrieval results in similar abundances of $\log(\mathrm{X_{H_2O}}){=}{-}5.39^{+0.42}_{-0.73}$, $\log(\mathrm{X_{CrH}}){=}{-}8.19^{+0.75}_{-0.77}$, and $\log(\mathrm{X_{K}}){=}{-}7.92^{+0.82}_{-1.59}$ (see the second row of Table \ref{table:retrievalcomparison}). This shows that including the other chemical species is not essential when retrieving abundances from this spectrum of WASP-31b. Increasing the number of opacity sources leads to a larger error on the retrieved parameters. This is as expected: Widening the allowed parameter space increases the number of possible solutions in the retrieval procedure.

\begin{table}
\caption{Retrieved atmospheric parameters using \textsc{TauREx} for the model shown in Figure \ref{fig:w31bspec}. The extended posterior distributions can be found in Figure \ref{fig:wasp31bpost} of the appendix.}             
\centering          
\begin{tabular}{l c}     
\hline\hline   
Parameter & Retrieved value \vspace{1mm}\\ \hline 
T$_{\mathrm{atm}}\,(\mathrm{K})$ & $1481^{+264}_{-355}$ \Tstrut \vspace{1mm} \\  
$\mathrm{R_{pl}}\,(\mathrm{R_J})$ & $1.48^{+0.02}_{-0.01}$ \vspace{1mm} \\  
$\log(\mathrm{P_{clouds}})\,(\mathrm{Pa})$ & $3.87^{+0.20}_{-0.20}$ \vspace{1mm} \\  
$\log(\mathrm{X_{H_2O}})$ & $-5.40^{+0.37}_{-0.43}$ \vspace{1mm} \\  
$\log(\mathrm{X_{CrH}})$ & $-8.51^{+0.62}_{-0.60}$ \vspace{1mm} \\  
$\log(\mathrm{X_{K}})$ & $-7.59^{+0.66}_{-0.94}$ \vspace{1mm} \\  

\hline                  
\end{tabular}
\label{table:retrievedparams}
\end{table}

\subsection{Comparison to previous work}
Earlier investigations of the transmission spectrum of WASP-31b found signatures of K, a grey cloud deck, and Rayleigh scattering \citep{sing2015, sing2016}. On top of that, a weak water absorption feature was found, which was also reported by later studies that used the same spectrum \citep{barstowetal2017, tsiaras2018}. \citet{tsiaras2018} found weak evidence for a water VMR of $\log(\mathrm{X_{H_2O}}){=}-3.84\pm1.90$. Other retrieved water abundances are equal to $-3.97^{+1.01}_{-2.27}$ \citep{pinhasetal2019} and $-4.55^{+1.77}_{-4.33}$ \citep{welbanksetal2019} with confidence levels of ${\sim}2\sigma$ and ${\sim}2.2\sigma$, respectively, and an abundance of $-3.27^{+1.44}_{-2.18}$ is found using the classical retrieval method within the ARCiS code \citep{minetal2020}. Our retrieved water abundance (see Table \ref{table:retrievedparams}) is lower but falls inside the error bars of the other investigations. \citet{sing2015} reported a $4.2\sigma$ significant detection of K, but the fidelity of this data point was later questioned by individual searches using the ground-based facilities FORS2 and UVES \citep{gibson2017, gibson2019} and IMACS \citep{mcgruderetal2020}. Our detection agrees with the $2.2\sigma$ detection that was made using combined FORS2 and STIS optical data \citep{gibson2017}. \citet{macdonaldetal2017} also found a weak detection ($2.2\sigma$) of NH$_3$, which was found at a similar abundance by \citet{minetal2020}. As can be seen in Table \ref{table:wasp31bresults}, the addition of NH$_3$ to our H$_2$O+CrH model also leads to an increase in the evidence level (DS${=}0.98$). As shown in Table \ref{table:jeffreysscale}, this was not seen as significant in our analysis (just below $2.1\sigma$). This small difference can be explained by the overlap in the CrH and NH$_3$ features around $1.5\,\mu\mathrm{m}$. Covering the features at longer wavelengths (e.g. at ${\sim}2.2\,\mu\mathrm{m}$) will greatly improve our ability to detect NH$_3$, as illustrated by \citet{macdonaldetal2017}.

Retrieved atmospheric temperatures for WASP-31b vary from $738^{+345}_{-231}\,\mathrm{K}$ \citep{minetal2020} to $1088.35\pm220.16\,\mathrm{K}$ \citep{tsiaras2018} and $1043^{+267}_{-172}\,\mathrm{K}$ \citep{pinhasetal2019}, all of which are exceeded by our retrieved temperature. Moreover, \citet{macdonaldetal2020} showed that hot Jupiter temperatures are generally underestimated by 1D retrievals. An explanation for this discrepancy in temperature might be an insufficient cloud model since \citet{sing2016} reported that the spectrum of this planet is not well explained by a single cloud model. A more complex model is included in \textsc{TauREx}, which parametrises the opacity caused by particle scattering following the Mie theory \citep{lee2013}. Adding this parametrisation to our best-fitting model also results in a relatively high $\mathrm{T_{atm}}{=}1507^{+239}_{-308}\,\mathrm{K}$ and low $\log(\mathrm{X_{H_2O}}){=}-5.73^{+0.51}_{-3.77}$ and is not found to be statistically significant. It is possible that this parametrisation is still not sufficiently complex. Another explanation may be the difference in opacity sources that are included: Our retrieval only includes H$_2$O, CrH, and K, whereas the others generally include H$_2$O, CH$_4$, CO, CO$_2$, and NH$_3$. Additionally, Na and K \citep{pinhasetal2019, minetal2020} and HCN \citep{pinhasetal2019} were also included in the retrievals. Increasing the number of opacity sources does indeed lead to a lower temperature of $\mathrm{T_{atm}}{=}1172^{+435}_{-226}\,\mathrm{K}$ (see the second row of Table \ref{table:retrievalcomparison}), consistent with the majority of earlier findings. With a Bayesian evidence level of 407.61, the addition of these opacity sources is not found to be statistically significant.

Associated with the lower temperature of this retrieval is an increase in the retrieved radius to $1.50\,\mathrm{R_J}$. As shown in Table \ref{table:retrievedparams}, the radius of WASP-31b is equal to $1.48\,\mathrm{R_J}$ in our best-fit model. This degeneracy between $\mathrm{R_{pl}}$ and T can also be seen from the posterior distributions in Figure \ref{fig:wasp31bpost}: Specifically, the plot in row 3 (from the bottom) and column 4 (from the left) shows that an increase in radius is degenerate with a decrease in temperature (due to its influence on the scale height). This degeneracy is another explanation for the difference with \citet{minetal2020}, who retrieve $\mathrm{R_{pl}}{=}1.51^{+0.02}_{-0.03}\,\mathrm{R_J}$. To test this suspicion, a retrieval was conducted assuming a fixed radius of $\mathrm{R_{pl}}{=}1.549$, and it did indeed result in a lower temperature of $1267^{+351}_{-288}\,\mathrm{K}$. However, this result is accompanied by unphysically high abundances of CrH and K (${\log(\mathrm{X})}{=}{-}3.61$ and ${-}0.18$, respectively). Using equilibrium temperatures, we can predict the maximum temperature of the terminator region to be ${\sim}1550\,\mathrm{K}$ for full redistribution ($f{=}1$) and perfect absorption of radiation ($A{=}0$). Hence, our retrieved temperature would be reasonable. The degeneracy that exists between temperature, radius, and abundances (e.g. \citealt{griffith2014, hengkitzmann2017}) may offer an explanation for the relatively low water abundances that we retrieve. A higher temperature leads to an overall increase in the scale height and thus the transit depth, dampening absorption features and leading to lower chemical abundances in the retrieval. 

Previous identifications of the signatures of CrH have been reported for brown dwarfs (e.g. \citealt{kirkpatricketal1999, kirkpatrickfull1999}). Specifically, \citet{02BuRaBe.CrH} found an abundance of $\mathrm{CrH/H_2}{\sim}2{-}4{\times}10^{-9}$ for the L5 dwarf 2MASSI J1507038-151648, which is in excellent agreement with the abundance that we retrieve for WASP-31b. \citet{macdonald2019} report a $4.1{\sigma}$ detection of metal hydrides in the transmission spectrum of exo-Neptune HAT-P-26b, and they identify three possible candidates: TiH ($4.1{\sigma}$), CrH ($2.1{\sigma}$), or ScH ($1.8{\sigma}$). As a possible candidate, CrH is retrieved at an abundance of  $-5.72^{+0.89}_{-1.37}$, which exceeds our value by almost three orders of magnitude. They propose vertical transport or secular contaminations by planetesimals as possible explanations for this high abundance and the fact that Cr is expected to have condensed out at the temperature of HAT-P-26b ($\mathrm{T_{eq}{\sim}1000}\,\mathrm{K}$).

The retrieved abundances can also be compared to the predictions from equilibrium chemistry, for example by using the GGChem code \citep{woitkehelling2018}. Around our retrieved temperature of WASP-31b, GGChem predicts CrH to be present at $\log(\mathrm{X_{CrH}}){\sim}{-}9$ for $\mathrm{P}{=}1\,\mathrm{bar}$ and solar composition, with lower abundances for lower pressures \citep{woitkehelling2018}. Hence, the retrieved CrH abundance of $-8.51^{+0.62}_{-0.60}$ is higher than predicted but still consistent. H$_2$O is expected at $\log(\mathrm{X_{H_2O}}){\sim}{-}3.3$ for the same temperature, about 100 times higher than the retrieved abundance. As previously stated, this might be related to degeneracies between different retrieval parameters. The fact that this large difference is not retrieved for the CrH abundance might also hint at an actual depletion of H$_2$O. Further observations can help in disclosing this.

\subsection{Chemistry}
From the first-row transition metals, Cr is the third most abundant element after Fe and Ni in the Sun \citep{asplundetal2009}. At the temperatures of close-in exoplanets, CrH is predicted to be an important Cr-bearing species. However, gaseous atomic Cr is expected to be the main Cr bearer, whereas significant fractions are also expected to be present in CrO or CrS \citep{woitkehelling2018}. These calculations were made assuming solar abundances and the corresponding solar abundance ratio $\log(\mathrm{Cr/O}){=}-3.05$ \citep{asplundetal2009}. If we make the simplifying assumption that for WASP-31b most of the Cr is in CrH, the planetary $\mathrm{Cr/O}$ abundance ratio can be calculated. At the temperature of WASP-31b, about half of the oxygen is expected in H$_2$O and the other half in CO \citep{madhusudhan2012, woitkehelling2018}. To correct for this, the retrieved H$_2$O abundance is multiplied by two, leading to a ratio of $\log(\mathrm{Cr/O}){=}{-}3.41$. The abundance ratio is lower than the solar value, but additional Cr is probably present in other species. Including the opacity data of these species in retrievals can lead to better constraints on the ratios, and the detectability of atomic Cr has recently been shown by its signatures in the ultra-hot Jupiter WASP-121b \citep{benyamietal2020}. Of course, abundance ratios may differ per star. For WASP-31, ratios are measured to be $\mathrm{O/H}{=}{+}0.06\,\mathrm{dex}$ and $\mathrm{Cr/H}{=}{-}0.08\,\mathrm{dex}$, relative to the Sun \citep{brewer2016}. This leads to a lower stellar abundance ratio of $\log(\mathrm{Cr/O}){=}{-}3.19$ for WASP-31, which may also partly explain the lower planetary ratio.

Monatomic Cr, the major gas-phase bearer at a wide range of temperatures ($300-3000$ K), is a refractory species. At the relevant pressures, it condenses into Cr metal between $1400$ and $1520$ K and into Cr$_2$O$_3$ at lower pressures of ${\sim}10^{-3}$ bar (e.g. \citealt{burrowssharp1999, loddersfegley2006, morleyetal2012}). The CrH abundance is related to the monatomic gas according to the equilibrium \citep{loddersfegley2006}:
\begin{align}
\ce{2 Cr (g) + H$_2$ <=> 2 CrH (g).}
\end{align}  
The condensation of Cr metal reduces the abundances of monatomic Cr and CrH, depleting the species from the atmosphere. Vertical mixing from lower, hotter layers is unlikely since Cr destruction reactions are highly exothermic at $\mathrm{T}{<}1400\,\mathrm{K}$, resulting in chemical lifetimes much shorter than the timescales of vertical mixing \citep{loddersfegley2006}. For the WASP-31b $\mathrm{T_{atm}}{=}1481^{+264}_{-355}\,\mathrm{K}$, the appearance of CrH would be reasonable since the gaseous Cr would not yet be fully depleted.

The discovery of Cr bearers can have implications for cloud formation in exoplanet atmospheres since they may cause the formation of Cr[s] clouds \citep{loddersfegley2006, morleyetal2012}. Moreover, \citet{leeblecichelling2018} suggested the possibility of  Cr[s] being seed particles that provide condensation surfaces for other cloud layers (e.g. sulphide or KCl cloud layers).

Lastly, the finding of Cr-bearing species in an atmosphere may give clues about the formation conditions of a planet. Because Cr is a refractory species, it is expected to be in the solid phase throughout most of the protoplanetary disk (e.g. \citealt{lodders2010}). Consequently, its presence on an exoplanet hints at the accretion of solid material during its formation. Determining the planetary Cr abundance (also in other Cr bearers) can then provide clues about the amount of solid accretion.

\subsection{Other observations}
Since the fidelity of the data point responsible for the K detection has been questioned \citep{gibson2017, gibson2019}, a few retrievals were conducted excluding the observed transit depth at ${\sim}0.77\,\mu\mathrm{m}$. CrH has absorption features around this wavelength, and these retrievals were done to make sure that the tentative CrH detection is not based on a disputed observation. A similar approach to what was described in Section \ref{sec:methodology} was followed. The resulting Bayesian evidence levels can be seen in Table \ref{table:nonkresults} and agree with our earlier findings, albeit with slightly lower DSs: The inclusion of both H$_2$O and CrH is preferred, with a DS of $7.97$ (or ${\sim}4.3\sigma$ confidence) over the flat model, or $2.98$ (${\sim}2.9\sigma$) over a model of only Rayleigh scattering and CIA. The influence of the measured transit depth at ${\sim}0.77\,\mu\mathrm{m}$ is explained by this decrease in DS, but, even without the measurement, statistically significant evidence for CrH is still found. In this case, a slightly lower $\mathrm{T_{atm}}{=}1339^{+332}_{-321}\,\mathrm{K}$ and higher $\mathrm{R_{pl}}{=}1.49^{+0.02}_{-0.02}\,\mathrm{R_J}$ are retrieved.

\begin{table}
\caption{Resulting Bayesian evidence levels and DSs for the \citet{sing2015} spectrum without $0.77\,\mu\mathrm{m}$ observation.}             
\label{table:nonkresults}      
\centering          
\begin{tabular}{l c c}     
\hline\hline                   
\textbf{Model parameters }& \textbf{log(E) }&\textbf{ DS }\vspace{1mm} \\ \hline 
Flat model      &  400.44   &   \\
Rayleigh+CIA    &  405.43   &  4.99 \vspace{1mm}    \\   \hline
Compared to flat model & & \\ \hline
CrH         &  406.01   &  5.58    \\
CrH+H$_2$O  &  408.40   &  7.97    \\  
\hline
\end{tabular}
\end{table}

\begin{figure}
\includegraphics[width=1\linewidth]{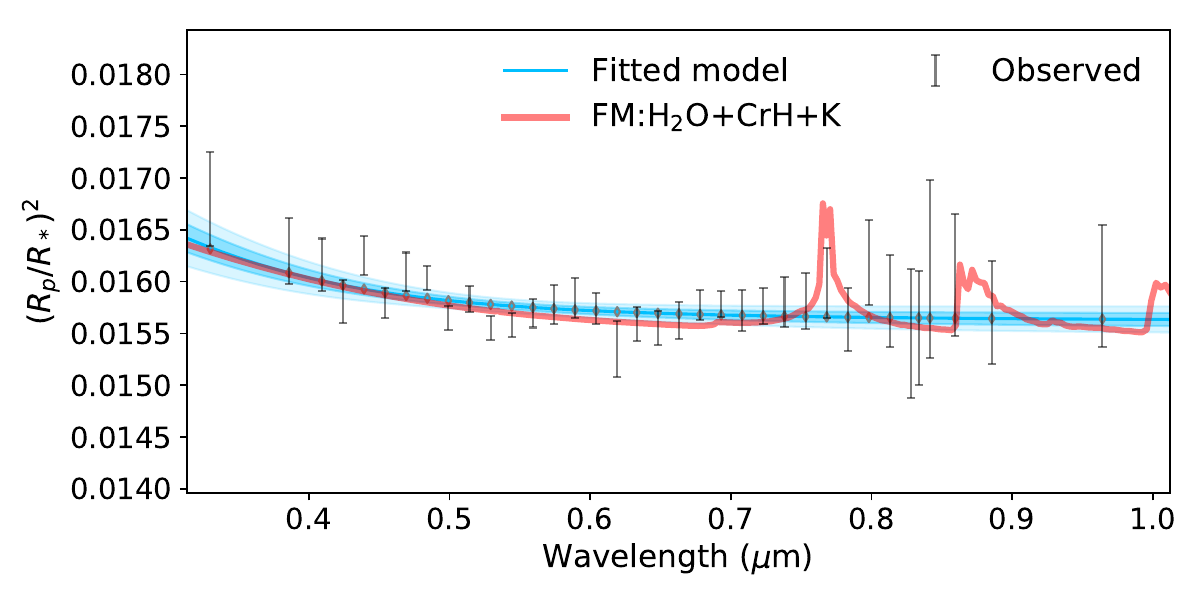}
\caption{Transmission spectrum as observed by FORS2/STIS with the best-fitting atmospheric model. The vertical error bars indicate the observed transit depths, and the different shadings in the upper panel represent $1$ and $2{\sigma}$ regions. A forward model based on the retrieval results in Table \ref{table:retrievedparams} is shown by the red line.}
\label{fig:w31bfors2spec}
\end{figure}

Ground-based optical data by the FORS2 at the VLT are also available for this planet \citep{gibson2017} and offer coverage from $0.4$ to $0.84\,\mu\mathrm{m}$. The combined FORS2/STIS data that are presented by \citet{gibson2017} were also analysed using \textsc{TauREx}.  With the same general setup, we retrieved the spectrum assuming five different models to test whether the CrH features are also found in the ground-based data. The resulting evidence levels and accompanying DSs are shown in Table \ref{table:fors2results}. Using these data, a model containing CrH is not found to be statistically significant over a model only containing Rayleigh scattering and CIA. Hence, in this case, the best-fitting atmospheric model was found to consist only of Rayleigh scattering and CIA and to be without any chemical species, indicating a cloudy atmosphere at a retrieved temperature of $\mathrm{T_{atm}}{=}1503^{+267}_{-369}\,\mathrm{K}$. The retrieved planetary radius agrees with the value we previously found at $1.48^{+0.02}_{-0.01}\,\mathrm{R_J}$. The spectrum and its lack of absorption features can be seen in Figure \ref{fig:w31bfors2spec}. While the measured transit depths between $0.7$ and $1.0\,\mu\mathrm{m}$ seem to show some signatures, they are not found to be statistically significant. For comparison, the forward model based on the retrieved parameters in Table \ref{table:retrievedparams} is shown by the red line, indicating the signatures that should be visible when CrH is present in the atmosphere. 

\begin{table}
\caption{Resulting Bayesian evidence levels and DSs for the combined FORS2/STIS data.}             
\label{table:fors2results}      
\centering          
\begin{tabular}{l c c}     
\hline\hline                   
\textbf{Model parameters }& \textbf{log(E) }&\textbf{ DS }\vspace{1mm} \\ \hline 
Flat model      &  241.49   &   \\
Rayleigh+CIA    &  245.36   &  3.87 \vspace{1mm}    \\   \hline
Compared to flat model & & \\ \hline
CrH+K       &  244.56   &  3.08    \\ 
CrH         &  244.72   &  3.24    \\
CrH+H$_2$O  &  245.50   &  4.01    \\  
\hline
\end{tabular}
\end{table}

Taking another look at the bottom image of Figure \ref{fig:w31bspecnospitz}, the orange line shows that the presence of CrH results in six prominent absorption peaks. Only three of these peaks (at $0.69$, $0.77$, and $0.88\,\mu\mathrm{m}$) fall (partially) inside the range probed by this combined FORS2/STIS spectrum, which can also be seen from the red line in Figure \ref{fig:w31bfors2spec}. To test whether the peaks in the WFC3 regime (at $1.18$ and $1.43\,\mu\mathrm{m}$) are driving the evidence for CrH, we added the WFC3 data to the combined FORS2/STIS data and conducted some additional retrievals on this spectrum. From the resulting evidence levels, as shown in Table \ref{table:fors2stiswfcresults}, it can be seen that this does not lead to a significant detection of CrH and only results in a preference for H$_2$O at ${\sim}3.7\sigma$ over a flat model. Using this spectral coverage, the temperature is retrieved to be $\mathrm{T_{atm}}{=}1476^{+277}_{-376}\,\mathrm{K}$ and the planetary radius agrees with our earlier findings at $1.48^{+0.02}_{-0.02}\,\mathrm{R_J}$. We can conclude that the evidence for CrH is driven by the observations made with STIS, specifically the observed transit depths around $0.88$ and $0.77\,\mu\mathrm{m}$. However, compatibility between different instruments cannot be taken for granted, and caution should thus be exercised when combining data, as illustrated by \citet{houyipetal2020} for the case of WASP-96b. This uncertainty, the non-detection when including ground-based data, and the broad continuous wavelength coverage that will be offered by future facilities further motivate the characterisation of WASP-31b.
\begin{table}
\caption{Resulting Bayesian evidence levels and DSs for a combined FORS2/STIS and WFC3 spectrum.}
\label{table:fors2stiswfcresults}      
\centering          
\begin{tabular}{l c c}     
\hline\hline                   
\textbf{Model parameters }& \textbf{log(E) }&\textbf{ DS }\vspace{1mm} \\ \hline 
Flat model      &  408.37   &   \\
Rayleigh+CIA    &  412.69   &  4.32 \vspace{1mm}    \\   \hline
Compared to flat model & & \\ \hline
FeH         &  410.81   &  2.45    \\ 
CrH         &  411.60   &  3.23    \\
H$_2$O      &  413.91   &  5.55   \vspace{1mm}    \\   \hline
Compared to H$_2$O-only model & & \\ \hline
H$_2$O + FeH    &  412.60   &  -1.31    \\ 
H$_2$O + AlH    &  412.86   &  -1.05    \\
H$_2$O + CP     &  413.23   &  -0.68    \\
H$_2$O + CrH    &  413.29   &  -0.62    \\
H$_2$O + ScH    &  413.37   &  -0.55    \\  
H$_2$O + NH$_3$ &  414.19   &   0.28    \\
\hline
\end{tabular}
\end{table}

\subsection{Near-future coverage}
\begin{figure}
\includegraphics[width=1.05\linewidth]{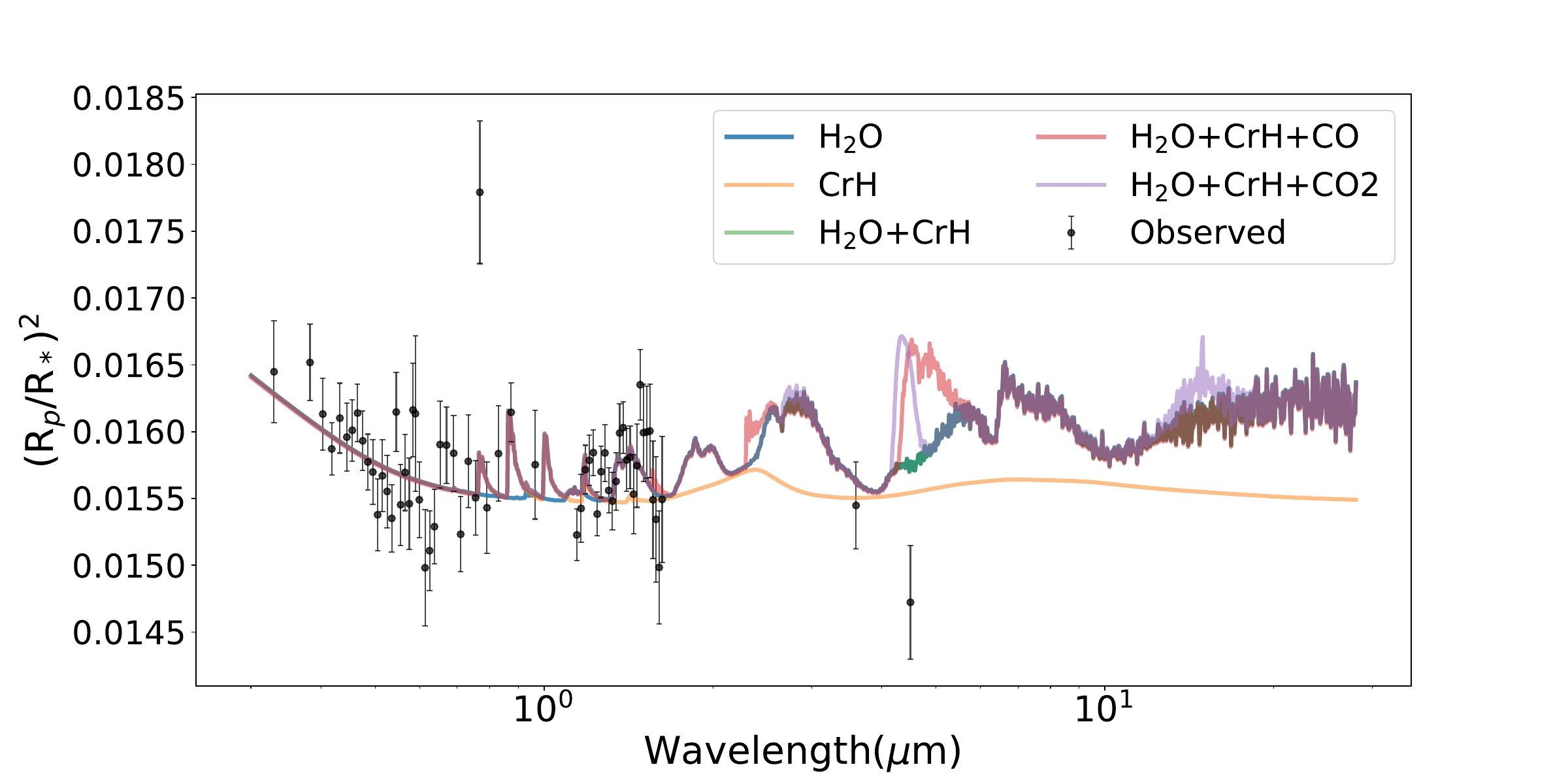}
\caption{Forward models for the transmission spectrum of WASP-31b in the spectral regime of JWST, based on the retrieved values for atmospheric parameters in Table \ref{table:retrievedparams}.}
\label{fig:w31bjwstrange}
\end{figure}

Figure \ref{fig:w31bjwstrange} shows the transmission spectrum for the wider wavelength range offered by the James Webb Space Telescope (JWST), scheduled for launch in October 2021. Using a variety of instruments, JWST will offer a spectral range from $0.6$ to $28\,\mu\mathrm{m}$ \citep{beichman2014}. The spectra in Figure \ref{fig:w31bjwstrange} correspond to different atmospheric compositions and are based on forward models that assume the retrieved best-fit values for atmospheric parameters; because of the discrepancy in the K detection, the alkali metal is not included in the models. 

The blue model is the only forward model that does not include CrH, and, by comparing it with the other models, it can be seen that the presence of CrH is purely based on the absorption signatures between $0.69$ to $1.43\,\mu\mathrm{m}$. Next to that, the presence of water can clearly be inferred from the familiar feature around $1.4\,\mu\mathrm{m}$ as well as the feature at $1.9\,\mu\mathrm{m}$, showing a clear distinction from the CrH-only model. Although many additional water signatures can be found at longer wavelengths, the fact that JWST's Near Infrared Imager and Slitless Spectrograph (NIRISS) can provide simultaneous coverage at $\mathrm{R}{\sim}700$ from $1$ to $2.5\,\mu\mathrm{m}$ makes these H$_2$O features and the CrH features in this regime interesting prospects for further characterisation. Another important goal is to derive improved C/O ratios, and combined observations from NIRISS and the Near Infrared Camera (NIRCam) are expected to deliver this \citep{stevensonetal2016jwst}, providing coverage from $1$ to $5\,\mu\mathrm{m}$. This is also evident from Figure \ref{fig:w31bjwstrange}, where a variation in transit depth can be seen around wavelengths of $4.5\,\mu\mathrm{m}$, depending on whether or not CO or CO$_2$ are included in the forward model. Hence, WASP-31b would be an interesting target for further characterisation, whereas the presence of CrH (and other metal hydrides) in exoplanets of similar temperatures is expected to be detectable with JWST. Because CrH rapidly condenses out for lower temperatures and the spectral signatures of more refractory materials such as TiO and VO start taking over at higher temperatures, its presence is probably detectable for planets with temperatures ranging from ${\sim}1300{-}2000\,\mathrm{K}$.

\section{Conclusions}\label{sec:conclusion}
In this study, a re-analysis of publicly available transmission data of the hot exoplanet WASP-31b has been conducted using the \textsc{TauREx II} retrieval framework. Transmission data from STIS, WFC3, and Spitzer provide spectral coverage between $0.3$ and $4.5\,\mu\mathrm{m}$. Assuming the simplified atmospheric representation in \textsc{TauREx} and out of the models that were fitted in this analysis, it was found that the spectrum is best explained by a model containing H$_2$O, CrH, and K in addition to H$_2$, He, a grey cloud deck, and Rayleigh scattering. As compared to a flat model without any spectral features, the H$_2$O-only model is statistically preferred at ${\sim}3.7\sigma$ and a CrH-only model at ${\sim}3.9\sigma$. A model with both H$_2$O and CrH was found at ${\sim}4.4\sigma$ and ${\sim}3.2\sigma$ over the flat model and a CIA+Rayleigh scattering model, respectively. Hence, we report the first statistical evidence for the signatures of CrH in an exoplanet atmosphere. Weak evidence for the addition of K to the atmospheric model was found at  ${\sim}2.2\sigma$ confidence over the H$_2$O+CrH model. As compared to earlier studies of WASP-31b, a relatively high temperature was retrieved, which can be explained by a combined influence of fewer opacity sources and degeneracies between temperature, radius, and chemical abundances.

The evidence for CrH naturally follows from its presence in brown dwarfs and is expected to be limited to planets with temperatures between $1300$ and $2000\,\mathrm{K}$. Cr-bearing species may play a role in the formation of clouds in exoplanet atmospheres, and their detection is also an indication of the accretion of solids during the formation of a planet. 

In additional retrievals, the disputed data point at ${\sim}0.77\,\mu\mathrm{m}$ was excluded, but evidence for the H$_2$O+CrH model was still found at a confidence level of ${\sim}4.3\sigma$ over the flat model. A combined FORS2/STIS spectrum was also available and tests were performed to confirm the CrH detection, but in this case no statistically significant CrH feature was found. By analysing the retrieval outcomes for different combinations of spectral coverage, it was found that the evidence for CrH is mostly based on the observed transit depths around $0.77$ and $0.88\,\mu\mathrm{m}$. Inspired by the non-agreement between different instruments, and using the best-fit atmospheric model for WASP-31b, it was shown that the spectral regime of JWST has the potential to confirm the CrH features.

\begin{acknowledgements}
We kindly thank Joanna Barstow for an excellent review which was valuable in improving the manuscript. We greatly appreciate the developers of TauREx (I. P. Waldmann, Q. Changeat, A. F. Al-Refaie, G. Tinetti, M. Rocchetto, E. J. Barton, S. N. Yurchenko and J. Tennyson), for making the \textsc{TauREx II} retrieval framework available and for their help in any inquiries. We would like to express our gratitude to the team who led the observations and made the data which was used in this work available (PI: D. K. Sing)\footnote{See \href{https://pages.jh.edu/\~dsing3/David\_Sing/Spectral\_Library.html}{https://pages.jh.edu/\(\sim\)dsing3/David\_Sing/Spectral\_Library.html} and \href{https://stellarplanet.org/science/exoplanet-transmission-spectra/}{https://stellarplanet.org/science/exoplanet-transmission-spectra/}.}.
\end{acknowledgements}

\bibliographystyle{aa} 
\bibliography{w31b.bib}

\begin{appendix} 
\section{Retrieval outputs}
\begin{minipage}{\textwidth}
    \includegraphics[width=\hsize]{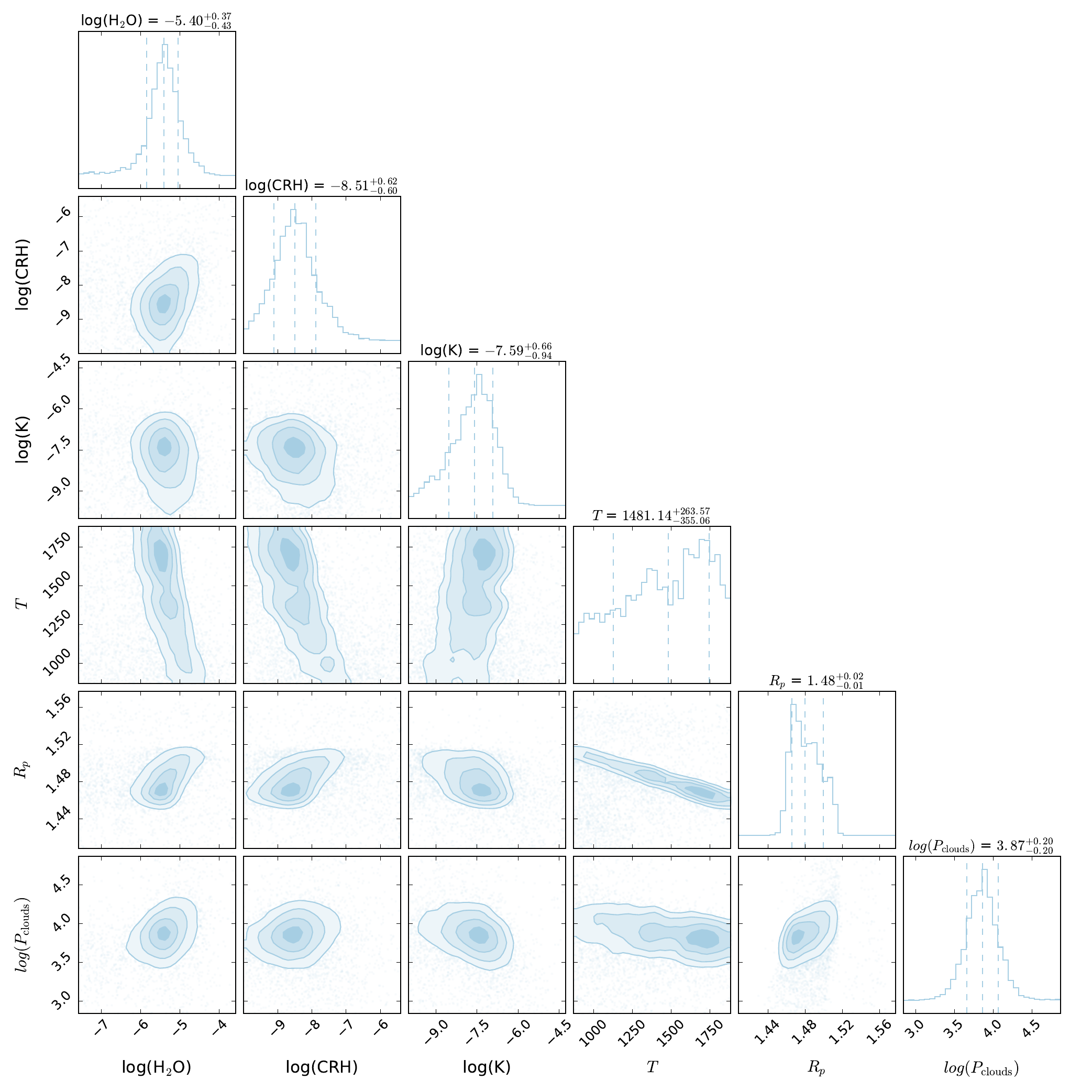}
    \captionof{figure}{\textsc{TauREx} posterior distributions of highest evidence retrieval for WASP-31b.}
    \label{fig:wasp31bpost}
\end{minipage}

\begin{table*}
\caption{Retrieved atmospheric parameters using \textsc{TauREx} for different retrieval setups and spectral coverage. We note that, besides the setups described here, the models also include H$_2$, He, a grey cloud deck, and Rayleigh scattering.}
\centering          
\begin{tabular}{l c c c c c c}     
\hline\hline   
Coverage and retrieval setup & T$_{\mathrm{atm}}\,(\mathrm{K})$ & $\mathrm{R_{pl}}\,(\mathrm{R_J})$ & $\log(\mathrm{P_{clouds}})\,(\mathrm{Pa})$ &
$\log(\mathrm{X_{H_2O}})$ & $\log(\mathrm{X_{CrH}})$ & $\log(\mathrm{X_{K}})$ \vspace{1mm}\\ \hline 
STIS + WFC3 + Spitzer & & & & & & \\ \cline{1-1}
H$_2$O+CrH+K (highest Evidence model) & $1481^{+264}_{-355}$  & $1.48^{+0.02}_{-0.01}$  & $3.87^{+0.20}_{-0.20}$  & $-5.40^{+0.37}_{-0.43}$  & $-8.51^{+0.62}_{-0.60}$  & $-7.59^{+0.66}_{-0.94}$ \Tstrut \vspace{1mm} \\  
H$_2$O+CrH+K+CO$_2$+CO+CH$_4$+NH$_3$+Na & $1172^{+435}_{-226}$  & $1.50^{+0.01}_{-0.02}$  & $4.03^{+1.07}_{-0.29}$  & $-5.39^{+0.42}_{-0.73}$  & $-8.19^{+0.75}_{-0.77}$  & $-7.92^{+0.82}_{-1.59}$ \Tstrut \vspace{1mm} \\ 
H$_2$O+CrH+K (Complex cloud) & $1507^{+239}_{-308}$  & $1.48^{+0.02}_{-0.01}$  & $3.86^{+0.18}_{-0.19}$  & $-5.41^{+0.33}_{-0.37}$  & $-8.53^{+0.57}_{-0.58}$  & $-7.63^{+0.67}_{-0.87}$ \Tstrut \vspace{1mm} \\ 
H$_2$O+CrH+K (Radius fixed at 1.549$\,\mathrm{R_J})$ & $1267^{+351}_{-288}$  & $1.549^{+0.00}_{-0.00}$  & $0.93^{+0.98}_{-1.27}$  & $-5.25^{+3.46}_{-3.20}$  & $-3.61^{+2.40}_{-3.26}$  & $-0.18^{+0.11}_{-0.18}$ \Tstrut \vspace{1mm} \\ 
H$_2$O+CrH & $1339^{+332}_{-321}$  & $1.49^{+0.02}_{-0.02}$  & $3.89^{+0.22}_{-0.21}$  & $-5.33^{+0.44}_{-0.44}$  & $-8.45^{+0.66}_{-0.63}$  & n/a \Tstrut \vspace{1mm} \\  \hline
STIS + WFC3 & & & & & & \\ \cline{1-1}
H$_2$O+CrH+K & $1614^{+184}_{-309}$  & $1.47^{+0.02}_{-0.01}$  & $3.86^{+0.18}_{-0.19}$  & $-5.35^{+0.29}_{-0.28}$  & $-8.65^{+0.56}_{-0.56}$  & $-7.54^{+0.64}_{-0.84}$ \Tstrut \vspace{1mm} \\  \hline
FORS2/STIS & & & & & & \\ \cline{1-1}
No species & $1503^{+267}_{-369}$  & $1.48^{+0.02}_{-0.01}$  & $3.66^{+0.18}_{-0.23}$  & n/a  & n/a  & n/a \Tstrut \vspace{1mm} \\ \hline
FORS2/STIS + WFC3 & & & & & & \\ \cline{1-1}
H$_2$O & $1476^{+277}_{-376}$  & $1.48^{+0.02}_{-0.01}$  & $3.76^{+0.17}_{-0.21}$  & $-5.51^{+0.38}_{-0.75}$  & n/a & n/a \Tstrut \vspace{1mm} \\
\hline                  
\end{tabular}
\label{table:retrievalcomparison}
\end{table*}
\end{appendix}
%
%

\end{document}